# Field-induced quantum critical point and nodal superconductivity in the heavy-fermion superconductor Ce$_2$PdIn$_8$

J. K. Dong[1], H. Zhang[1], X. Qiu[1], B. Y. Pan[1], Y. F. Dai[1], T. Y. Guan[1], S. Y. Zhou[1], D. Gnida[2], D. Kaczorowski[2] & S. Y. Li[1*]

[1]*Department of Physics, State Key Laboratory of Surface Physics, and Laboratory of Advanced Materials, Fudan University, Shanghai 200433, China*

[2]*Institute of Low Temperature and Structure Research, Polish Academy of Sciences, P. O. Box 1410, 50-950 Wroclaw, Poland*

**The interplay between magnetism and superconductivity has been a central issue in unconventional superconductors. While the dynamic magnetism could be the source of electron pairing, the static magnetism is generally believed to compete with superconductivity. In this sense, the observation of *Q* phase, the coupled spin-density wave order and superconductivity, in the heavy-fermion superconductor CeCoIn$_5$ is very puzzling. Whether this *Q* phase origins from the novel Fulde-Ferrel-Larkin-Ovchinnikov state is under hot debate. Here we report the resistivity and thermal conductivity study of a newly discovered heavy-fermion superconductor Ce$_2$PdIn$_8$ down to 50 mK. We find an unusual field-induced quantum critical point at the upper critical field $H_{c2}$ and unconventional nodal superconductivity in Ce$_2$PdIn$_8$. The jump of thermal conductivity κ(*H*)/*T* near $H_{c2}$ suggests a first-order-like phase transition at low temperatures. These results mimic the features of the *Q* phase in CeCoIn$_5$, implying that Ce$_2$PdIn$_8$ is another promising compound to investigate the exotic *Q* phase and FFLO state. The**

**comparison between $CeCoIn_5$ and $Ce_2PdIn_8$ may help to clarify the origin of the $Q$ phase.**

The unconventional superconductivity remains a major challenge in current condensed matter physics[1]. For many unconventional superconductors, the superconductivity emerges near a magnetic quantum critical point (QCP), which suggests that the dynamic magnetism (or spin fluctuations) may be the source of electron pairing[1]. On the contrary, static magnetism is generally believed to compete with superconductivity, as demonstrated in $La_{2-x}Sr_xCuO_4$ and $CeRhIn_5$ where static magnetism is induced by magnetic field and finally defeats the superconductivity in the normal state[2,3].

However, there is one exceptional case, the heavy-fermion superconductor $CeCoIn_5$ with $T_c$ = 2.3 K at ambient pressure[4]. Its superconducting gap has $d$-wave symmetry[5,6]. While there is no static magnetism in $CeCoIn_5$ at zero field, NMR, neutron scattering, and muon spin rotation (µSR) experiments have provided clear evidences for a field-induced magnetism in the low-temperature-high-field (LTHF) part of the phase diagram[7-12]. It was identified as an antiferromagnetic (AF) spin-density wave (SDW) order with an incommensurate modulation $\mathbf{Q}$ = (0.44, 0.44, 0.5)[8]. Interestingly enough, this SDW order disappears in the normal state above $H_{c2}$, showing that magnetic order and superconductivity in $CeCoIn_5$ are directly coupled[7,8].

While this has nicely explained the field-induced QCP at $H_{c2}$ (refs 13, 14), the physical origin of this LTHF superconducting $Q$ phase is still under debate. For example, Yanase and Sigrist suggested that the incommensurate SDW order is stabilized in the Fulde-Ferrell-Larkin-Ovchinnikov (FFLO) state by the appearance of the Andreev bound state localized around the zeros of the FFLO order parameter[15]. The novel FFLO state with broken spatial symmetry was predicted in 1960s[16,17], but has never been experimentally verified before. CeCoIn$_5$ has been considered as the strongest candidate for the formation of the FFLO state, and stimulated extensive studies[18]. However, Aperis, Varelogiannis, and Littlewood argued that the $Q$ phase is a pattern of coexisting condensates: a $d$-wave singlet superconducting state, a staggered $\pi$-triplet superconducting state, and an SDW[19]. In this latter scenario, the FFLO state is excluded. Therefore it is still unclear whether the $Q$ phase relates to an FFLO state.

To clarify this important issue, it will be very helpful to find more compounds with this kind of $Q$ phase. One natural way is to examine those related compounds, Ce$_n$MIn$_{3n+2}$ (M = Co, Rh, Ir). For CeRhIn$_5$, field-induced magnetism was found to coexist with superconductivity under a pressure of 1.77 - 2.25 GPa[3]. However, this magnetic order persists above $H_{c2}$, as in La$_{2-x}$Sr$_x$CuO$_4$ (ref. 2), thus does not couple to superconductivity. For CeIrIn$_5$ with bulk $T_c$ = 0.4 K, there is a possible metamagnetic QCP at $H_m$ = 28 T, which is well above $H_{c2}$ (ref. 20). So far, there were rare experiments to search for field-induced magnetism and QCP in Ce$_2$CoIn$_8$ ($T_c$ = 0.4 K)[21], Ce$_2$RhIn$_8$ ($T_c$ = 2 K under pressure)[22], and heavy-fermion paramagnet Ce$_2$IrIn$_8$ (ref. 23).

Recently, it was discovered that Ce$_2$PdIn$_8$ is also a heavy-fermion superconductor with $T_c$ = 0.68 K at ambient pressure[24]. Here we report the observations of a field-induced QCP, the nodal superconductivity, and a first-order-like phase transition at $H_{c2}$ in Ce$_2$PdIn$_8$ by resistivity and thermal conductivity measurements. Comparing with CeCoIn$_5$, our findings suggest that Ce$_2$PdIn$_8$ is another playground to investigate the exotic $Q$ phase and FFLO state.

**Results**

**Field evolution of resistivity**. Fig. 1(a) shows the in-plane resistivity of our Ce$_2$PdIn$_8$ single crystal in zero field. A broad peak is found at $T_{coh}$ = 25 K, which is attributed to a crossover between incoherent scattering at high temperature and the development of a coherent heavy-fermion state at lower temperature. The curve is very smooth below 20 K, suggesting that there are no impurity phases, especially the antiferromagnetic $CeIn_3$ with $T_N \sim 10$ K (ref. 24). In Fig. 1(b), the low-temperature $\rho(T)$ in magnetic fields $H \parallel c$ up to 11 T are plotted. From the zero-field data, $T_c = 0.68$ K is obtained, which is defined at the 10% of the normal-state value. The 10-90% transition width is only 20 mK. It is found that $\rho(T)$ obeys $T$-linear dependence nicely above $T_c = 0.68$ K, up to about 2 K. A linear fit of the data between 0.7 and 1.5 K gives a residual resistivity $\rho_0 = 2.41$ $\mu\Omega$ cm.

With increasing field, $T_c$ gradually decreases, to 50 mK at $H = 2.32$ T. This field is determined as the bulk $H_{c2}$ for $H \parallel c$. In slightly higher field $H = 2.4$ T, the resistive transition is completely suppressed, and the $T$-linear behaviour of normal-state $\rho(T)$ persists all the way down to 50 mK. In $H > 2.4$ T, The curves show clear deviation from the $T$-linear dependence. The data of $H = 2.4, 3, 4, 5, 6, 8$, and 11 T are plotted as $\rho$ vs $T^2$ in Fig. 1(c). It is clearly seen that a Fermi liquid behaviour of resistivity, $\rho \sim AT^2$, develops with increasing field.

**Phase diagram**. Based on these resistivity results, we have constructed an $H$ - $T$ phase diagram for $Ce_2PdIn_8$ in the main panel of Fig. 2. The field dependence of the coefficient $A(H)$ is plotted in the inset, which tends to diverge towards $H_{c2} = 2.32$ T. The fitting of $A(H) = A_0 (H - 2.32)^\alpha$ gives $\alpha = -0.57 \pm 0.02$. Note that for the critical behaviour in $A(H)$ near the QCP of $YbRh_2Si_2$ (ref. 25), $YbAlB_4$ (ref. 26), $CeCoIn_5$ ($\rho_{ab}$ in $H \parallel c$)[13] and $CeRhIn_5$ ($\rho_c$ in $H \parallel ab$ and under 2.35 GPa)[27], $\alpha = -1, -0.5, -1.37$ and $-0.5$ were obtained, respectively.

**Field evolution of thermal conductivity**. Fig. 3 shows the temperature dependence of the in-plane thermal conductivity in $H \parallel c$ up to 2.4 T, plotted as $\kappa/T$ vs $T$. The Wiedemann-Franz (WF) law is firstly checked in the normal state. The $H = 2.4$ T curve is fitted to $\kappa/T = 1/(a+bT)$, since $\rho(2.4T)$ is $T$-linear dependent and electrons dominate the heat transport at low temperature. The extrapolation gives $\kappa_0/T(2.4T) = 1/a \approx 9.28$ mW K$^{-2}$ cm$^{-1}$. This value is about 94% of the WF law expectation $L_0/\rho_0(2.4T) = 9.84$ mW K$^{-2}$ cm$^{-1}$, with $L_0 = 2.45 \times 10^{-8}$ W$\Omega$K$^{-2}$ and $\rho_0(2.4T) = 2.49$ $\mu\Omega$ cm. The rough satisfaction of WF law in the normal state shows that our thermal conductivity data are reliable.

In zero field, $\kappa/T$ behaves similarly to that of CeCoIn$_5$ (refs 28, 29), increasing below $T_c$, showing a broad peak at ~ 0.45 K, then decreasing towards $T = 0$. Below 250 mK, $\kappa/T \propto T$ and extrapolates to $\kappa_0/T = 2.09 \pm 0.02$ mW K$^{-2}$ cm$^{-1}$, more than 20% of the normal-state value. For the $d$-wave superconductor CeCoIn$_5$, one measurement down to 50 mK obtained $\kappa_0/T = 17$ mW K$^{-2}$ cm$^{-1}$ (ref. 28), while another measurement down to 10 mK got $\kappa_0/T \approx 3$ mW K$^{-2}$ cm$^{-1}$ (ref. 29). The significant $\kappa_0/T$ of Ce$_2$PdIn$_8$ in zero field is a strong evidence for nodes in the superconducting gap[30], which shows that Ce$_2$PdIn$_8$ is also an unconventional superconductor. This is not very surprising, since previously the $d$-wave superoncuducting gap has been demonstrated in CeCoIn$_5$ (refs 5, 6), CeRhIn$_5$ under pressure[31], and CeIrIn$_5$ (ref. 32).

**Field dependence of $\kappa/T$**. The field dependence of $\kappa_0/T$ may provide further support for the nodes[30]. In Fig. 3, however, the temperature dependence of $\kappa/T$ changes dramatically in fields, and it is hard to extrapolate $\kappa_0/T$ reliably for all the curves. Therefore we plot $\kappa/T$ at 60 mK, normalized to its normal-state value (in bulk $H_{c2} = 2.32$ T), vs $H/H_{c2}$ in Fig. 4. For comparison, similar data of the clean $s$-wave superconductor Nb (ref. 33) and an overdoped $d$-wave superconductor Tl-2201 (ref. 34) at $T \to 0$ are also shown. At low field ($H/H_{c2} < 0.5$), $\kappa(H)/T$ of Ce$_2$PdIn$_8$ shows downward curvature as in Tl-2201, providing further support for the nodes in the superconducting gap. In contrast to the $s$-wave superconductor Nb, the rapid increase of $\kappa(H)/T$ at low field in nodal superconductors results from the Volovik effect of the nodal quasiparticles[35].

It is worth to point out that the $\kappa(H)/T$ of $Ce_2PdIn_8$ at 60 mK is quit different from the abnormal one of $CeCoIn_5$ at 50 mK[29], which is also plotted in Fig. 4. For $CeCoIn_5$, the decrease of $\kappa/T$ above 20 mT was attributed to the vortex scattering, since the quasiparticle mean free path $l$ in zero field is very large ($\approx$ 4000 Å) due to the high purity of the sample[29]. However, for $Ce_2PdIn_8$, $l \approx$ 420 Å has been estimated[24], roughly one order smaller than $CeCoIn_5$. In this case, vortex scattering does not show significant effect, and $\kappa(H)/T$ of $Ce_2PdIn_8$ behaves more like the typical $d$-wave superconductor Tl-2201 at low field[34].

**Discussion**

From the phase diagram of $Ce_2PdIn_8$ in Fig. 2, the restoration of a Fermi liquid state above $H_{c2}$ and the critical behaviour of $A(H)$ provide clear evidences for a field-induced QCP at $H_{c2}$. This kind of QCP is very unusual and rare. It was first found in $CeCoIn_5$ by resistivity and specific heat measurements[13,14]. Initially, it was very puzzling why this QCP is located right at $H_{c2}$. The identification of the LTHF $Q$ phase, in which the AF SDW order simultaneously disappears with superconductivity at $H_{c2}$[7,8], clarifies that it is actually an AF SDW QCP. Recently, a field-induced QCP at $H_{c2}$ was also observed in $CeRhIn_5$ under optimal pressure $p_2 \approx$ 2.35 GPa (ref. 27). However, according to the phase diagram of $CeRhIn_5$ (ref. 3), this QCP separates the pure superconducting state with the field-induced magnetism in the normal state, therefore it is different from the one in $CeCoIn_5$. In our study of $Ce_2PdIn_8$, no pressure is applied, and the field-induced QCP is more like the case in $CeCoIn_5$. To confirm this, more experiments such as NMR, neutron scattering, and μSR are needed to probe the magnetic order below and above $H_{c2}$, as have been done in $CeCoIn_5$ (refs 7-12).

In Fig. 4, $\kappa(H)/T$ of CeCoIn$_5$ shows an abrupt jump near $H_{c2}$. This is an evidence for the first-order transition[10,36,37], which apparently corresponds to the disappearance of the $Q$ phase at $H_{c2}$ (refs 7, 8). For Ce$_2$PdIn$_8$, a less sharp jump of $\kappa(H)/T$ is also found near $H_{c2}$, which hints that the superconducting to normal state transition at very low temperature is first-order-like. This result, together with the unusual field-induced QCP at $H_{c2}$ and the nodal superconductivity, all mimic those features of the $Q$ phase in CeCoIn$_5$. Therefore, it will be very interesting to search for the exotic $Q$ phase in this second compound other than CeCoIn$_5$.

Since CeCoIn$_5$ is the strongest candidate for an FFLO state[18] and the $Q$ phase in it may origin from the FFLO state[15], we check the conditions for Ce$_2$PdIn$_8$ to form an FFLO state. There are two requirements for the formation of FFLO state: very large Maki parameter ($\alpha > 1.8$) and very clean compound ($\xi \ll l$)[18]. The Maki parameter $\alpha = \sqrt{2} H_{orb}/H_P$, where $H_{orb}$ and $H_P$ are the orbital and Pauli limiting $H_{c2}$, reflects the relative strength of Pauli paramagnetic effect and orbital effect when magnetic field is suppressing the superconductivity[38]. For CeCoIn$_5$, $\alpha = 4.6$ and 5.0 were obtained in $H \parallel ab$ and $H \parallel c$, respectively[39]. For our Ce$_2$PdIn$_8$ in $H \parallel c$, $H_{orb} = -0.7 T_c dH_{c2}/dT|_{T=T_c} \approx 6.8$ T is obtained from the initial slop $dH_{c2}/dT|_{T=T_c} = -14.3$ T/K in Fig. 2. By assuming $H_P \approx H_{c2}(0) = 2.32$ T, we get $\alpha \approx 2.9$. Therefore, both CeCoIn$_5$ and Ce$_2$PdIn$_8$ meet the required minimum value of $\alpha$. The in-plane superconducting coherence length $\xi = 47 - 50$ Å was estimated for CeCoIn$_5$ (ref. 40). This value is about 100 times smaller than the mean free path $l \approx 4000$ Å[29], showing that CeCoIn$_5$ is in the clean limit $\xi \ll l$. For Ce$_2$PdIn$_8$, $\xi \approx 82$ Å and $l \approx 420$ Å were estimated[24]. This shows that Ce$_2$PdIn$_8$ is a clean compound ($\xi \ll l$) too, although not as clean as CeCoIn$_5$. Therefore Ce$_2$PdIn$_8$ also fulfils the requirements for the formation of FFLO state.

To summarize, we have measured the resistivity and thermal conductivity of the heavy-fermion superconductor Ce$_2$PdIn$_8$ single crystal down to 50 mK. An unuaual field-induced QCP at $H_{c2}$ is demonstrated by the observation of $\rho \sim T$ near $H_{c2}$ and the development of $\rho \sim T^2$ Fermi liquid behavior at higher fields. The large $\kappa_0/T$ at zero field and the rapid increase of $\kappa(H)/T$ at low field give strong evidences for unconventional nodal superconductivity. Moreover, the jump of $\kappa(H)/T$ near $H_{c2}$ indicates a first-order-like phase transition from the superconducting to the normal state at low temperatures. Since these features are very similar to those of the $Q$ phase in CeCoIn$_5$, and Ce$_2$PdIn$_8$ also meets the requirements for the formation of FFLO state, we conclude that Ce$_2$PdIn$_8$ is another promising compound to investigate the exotic $Q$ phase and FFLO state. The comparison between CeCoIn$_5$ and Ce$_2$PdIn$_8$ may help to clarify the physical origin of the $Q$ phase.

**Methods**

**Sample preparation**. Single crystals of Ce$_2$PdIn$_8$ were grown from In-flux[24]. It can be viewed as periodic stacking of two layers of CeIn$_3$ on a layer of PdIn$_2$. To avoid the impurity phase of CeIn$_3$ (ref. 24), very thin crystals were selected for this study, with the typical thickness 40 μm.

**Resistivity and thermal conductivity measurements**. Four contacts were made with soldered indium, which were used for both resistivity and thermal conductivity measurements. The resulting contact resistance is ~ 5 mΩ at 1.5 K. In-plane thermal conductivity was measured in a dilution refrigerator, using a standard four-wire steady-state method with two RuO$_2$ chip thermometers, calibrated *in situ* against a reference RuO$_2$ thermometer. Magnetic fields were applied along the *c* axis and perpendicular to the heat current. To ensure a homogeneous field distribution in the sample, all fields were applied at temperature above $T_c$.

**References**

[1] Norman, M. R. The challenge of unconventional superconductivity. *Science* **332**, 196-200 (2011).

[2] Lake, B. *et al.* Antiferromagnetic order induced by an applied magnetic field in a high-temperature superconductor. *Nature* (London) **415**, 299-302 (2002).

[3] Park, T. *et al.* Hidden magnetism and quantum criticality in the heavy fermion superconductor $CeRhIn_5$. *Nature* (London) **440**, 65-68 (2006).

[4] Petrovic, C. *et al.* Heavy-fermion superconductivity in $CeCoIn_5$ at 2.3 K. *J. Phys.: Condens. Matter* **13**, L337-L342 (2001).

[5] Izawa, K. *et al.* Angular position of nodes in the superconducting gap of quasi-2D heavy-fermion superconductor $CeCoIn_5$. *Phys. Rev. Lett.* **87**, 057002 (2001).

[6] An, K. *et al.* Sign reversal of field-angle resolved heat capacity oscillations in a heavy fermion superconductor $CeCoIn_5$ and $d_{x^2-y^2}$ pairing symmetry. *Phys. Rev. Lett.* **104**, 037002 (2010).

[7] Young, B. L. *et al.* Microscopic evidence for field-induced magnetism in $CeCoIn_5$. *Phys. Rev. Lett.* **98**, 036402 (2007).

[8] Kenzelmann, M. *et al.* Coupled superconducting and magnetic order in $CeCoIn_5$. *Science* **321**, 1652-1654 (2008).

[9] Spehling, J. *et al.* Field-induced coupled superconductivity and spin density wave order in the heavy fermion compound $CeCoIn_5$. *Phys. Rev. Lett.* **103**, 237003 (2009).

[10] Koutroulakis, G. *et al.* Field evolution of coexisting superconducting and magnetic orders in $CeCoIn_5$. *Phys. Rev. Lett.* **104**, 087001 (2010).

[11] Kenzelmann, M. *et al.* Evidence for a magnetically driven superconducting Q phase of $CeCoIn_5$. *Phys. Rev. Lett.* **104**, 127001 (2010).

[12] Kumagai, K. *et al.* Evolution of paramagnetic quasiparticle excitation emerged in the high-field superconducting phase of CeCoIn$_5$. *Phys. Rev. Lett.* **106**, 137004 (2011).

[13] Paglione, J. *et al.* Field-induced quantum critical point in CeCoIn$_5$. *Phys. Rev. Lett.* **91**, 246405 (2003).

[14] Bianchi, A. *et al.* Avoided antiferromagnetic order and quantum critical point in CeCoIn$_5$. *Phys. Rev. Lett.* **91**, 257001 (2003).

[15] Yanase, Y. & Sigrist, M. Antiferromagnetic order and π–triplet pairing in the Fulde-Ferrel-Larkin-Ovchinnikov state. *J. Phys. Soc. Jpn.* **78**, 114715 (2009).

[16] Fulde, P. & Ferrel, R. A. Superconductivity in a strong spin-exchange field. *Phys. Rev.* **135**, A550-A563 (1964).

[17] Larkin, A. I. & Ovchinnikov, Y. N. Inhomogenous state of superconductors. *Zh. Eksp. Teor. Fiz.* **47**, 1136 (1964) [*Sov. Phys. JETP* **20**, 762 (1965)].

[18] Matsuda, Y. & Shimahara, H. Fulde-Ferrel- Larkin-Ovchinnikov state in heavy fermion superconductors. *J. Phys. Soc. Jpn.* **76**, 051005 (2007).

[19] Aperis, A., Varelogiannis, G., & Littlewood, P. B. Magnetic-field-induced pattern of coexisting condensates in the superconducting state of CeCoIn$_5$. *Phys. Rev. Lett.* **104**, 216403 (2010).

[20] Capan, C. *et al.* Unusual metamagnetism in CeIrIn$_5$. *Phy. Rev. B* **80**, 094518 (2009).

[21] Chen, G. *et al.* Observation of superconductivity in heavy-fermion compounds of Ce$_2$CoIn$_8$. *J. Phys. Soc. Jpn.* **71**, 2836-2838 (2002).

[22] Nicklas, M. *et al.* Magnetism and superconductivity in Ce$_2$RhIn$_8$. *Phys. Rev. B* **67**, 020506 (2003).

[23] Thompson, J. D. *et al.* Superconductivity and magnetism in a new class of heavy-fermion materials. *J. Magn. Magn. Mater.* **226-230**, 5-10 (2001).

[24] Kaczorowski, D. *et al.* Emergence of a superconducting state from an antiferromagnetic phase in single crystals of the heavy fermion compound $Ce_2PdIn_8$. *Phys. Rev. Lett.* **103**, 027003 (2009). Kaczorowski, D. *et al.* Reply. *Phys. Rev. Lett.* **104**, 059702 (2010).

[25] Gegenwart, P. *et al.* Magnetic-field induced quantum critical point in $YbRh_2Si_2$. *Phys. Rev. Lett.* **89**, 056402 (2002).

[26] Nakatsuji, S. *et al.* Superconductivity and quantum criticality in the heavy-fermion system $YbAlB_4$. *Nature Phys.* **4**, 603-607 (2008).

[27] Park, T. *et al.* Field-induced quantum critical point in the pressure-induced superconductor $CeRhIn_5$. *Phys. Status Solidi B* **247**, 553-556 (2010).

[28] Tanatar, M. A. *et al.* Unpaired electrons in the heavy-fermion superconductor $CeCoIn_5$. *Phys. Rev. Lett.* **95**, 067002 (2005).

[29] Seyfarth, G. *et al.* Multigap superconductivity in the heavy-fermion system $CeCoIn_5$. *Phys. Rev. Lett.* **101**, 046401 (2008).

[30] Shakeripour, H., Petrovic, C., & Taillefer, L. Heat transport as a probe of superconducting gap structure. *New J. Phys.* **11**, 055065 (2009).

[31] Park, T. *et al.* Probing the nodal gap in the pressure-induced heavy fermion superconductor $CeRhIn_5$. *Phys. Rev. Lett.* **101**, 177002 (2008).

[32] Kasahara, Y. *et al.* Thermal conductivity evidence for a $d_{x^2-y^2}$ pairing symmetry in the heavy-fermion $CeIrIn_5$. *Phys. Rev. Lett.* **100**, 207003 (2008).

[33] Lowell, J. & Sousa, J. B. Mixed-state thermal conductivity of type II superconductors. *J. Low. Temp. Phys.* **3**, 65-87 (1970).

[34] Proust, C. *et al.* Heat transport in a strongly overdoped cuprate: Fermi liquid and a pure *d*-wave BCS superconductor. *Phys. Rev. Lett.* **89**, 147003 (2002).

[35] Volovik, G. E. Superconductivity with lines of gap nodes – density-of-state in the vortex. *JETP Lett.* **58**, 469-473 (1993).

[36] Bianchi, A. *et al.* First-order superconducting phase transitions in CeCoIn$_5$. *Phys. Rev. Lett.* **89**, 137002 (2002).

[37] Bianchi, A. *et al.* Possible Fulde-Ferrell-Larkin-Ovchinnikov superconducting state in CeCoIn$_5$. *Phys. Rev. Lett.* **91**, 187004 (2003).

[38] Maki, K. Effect of Pauli paramagnetism on magnetic properties of high-field superconductors. *Phys. Rev.* **148**, 362-369 (1966).

[39] Kumagai, K. *et al.* Fulde-Ferrel- Larkin-Ovchinnikov state in a perpendicular field of quasi-two-dimensional CeCoIn$_5$. *Phys. Rev. Lett.* **97**, 227002 (2006).

[40] DeBeer-Schmitt, L. *et al.* Field-dependent coherent length in the superclean, high-κ superconductor CeCoIn$_5$. *Phys. Rev. Lett.* **97**, 127001 (2006).

**Acknowledgements** We thank W. Bao, Y. Chen, and Y. Y. Wang for discussions. This work is supported by the Natural Science Foundation of China, the Ministry of Science and Technology of China (National Basic Research Program No: 2009CB929203), Program for New Century Excellent Talents in University, Program for Professor of Special Appointment (Eastern Scholar) at Shanghai Institutions of Higher Learning, and STCSM of China (No: 08dj1400200 and 08PJ1402100).

**Author Information** Reprints and permissions information is available at www.nature.com/reprints. The authors declare no competing financial interests. Correspondence and requests for materials should be addressed to S. Y. Li (shiyan_li@fudan.edu.cn).

**Figure 1 | Field evolution of resistivity.**

**a,** In-plane resistivity of $Ce_2PdIn_8$ single crystal in zero field. The curve is very smooth below 20 K, showing no impurity phases in the sample. **b,** Low-temperature resistivity of $Ce_2PdIn_8$ in magnetic fields applied along *c* axis. The solid line is a linear fit of the zero-field data between 0.7 and 1.5 K. **c,** $\rho$ vs $T^2$ for $H$ = 2.4, 3, 4, 5, 6, 8, and 11 T (data sets are offset for clarity). The solid lines are fits to $\rho = \rho_0 + AT^2$. The arrows indicate the upper limit of the temperature range of $T^2$ behaviour.

**Figure 2 | Phase diagram.**

The field-temperature phase diagram of $Ce_2PdIn_8$ determined from the resistivity measurements. The $T_c$ is defined at the 10% of the normal-state resistivity in each field. The $T_{FL}$ is defined as the upper limit of the temperature range of $T^2$ dependence Fermi liquid behaviour. The inset shows the field dependence of the coefficient $A$ of $\rho = \rho_0 + AT^2$, which tends to diverge towards $H_{c2}$ = 2.32 T. This phase diagram suggests a field-induced quantum critical point located at $H_{c2}$ = 2.32 T.

**Figure 3 | Field evolution of thermal conductivity.**

Low-temperature in-plane thermal conductivity of $Ce_2PdIn_8$ in magnetic fields applied along the *c* axis. The saturation of $\kappa/T$ towards 2.32 T confirms it as the bulk $H_{c2}$. The solid lines are fits to the $H = 0$ and 2.4 T data (see text). The dashed line is the normal-state Wiedemann-Franz law expectation $L_0/\rho_0(2.4T)$, with $L_0 = 2.45 \times 10^{-8}$ W$\Omega$K$^{-2}$ and $\rho_0(2.4T) = 2.49$ $\mu\Omega$ cm.

**Figure 4 | Field dependence of $\kappa/T$.**

Normalized $\kappa/T$ to the normal-state value for $Ce_2PdIn_8$ at 60 mK, and for $CeCoIn_5$ at 50 mK (ref. 29), as a function of $H/H_{c2}$. For both compounds, the sharp jump of $\kappa/T$ near $H_{c2}$ indicates a first-order phase transition. Similar data of the clean *s*-wave superconductor Nb (ref. 33) and an overdoped *d*-wave superconductor Tl-2201 (ref. 34) at $T \to 0$ are also shown for comparison.

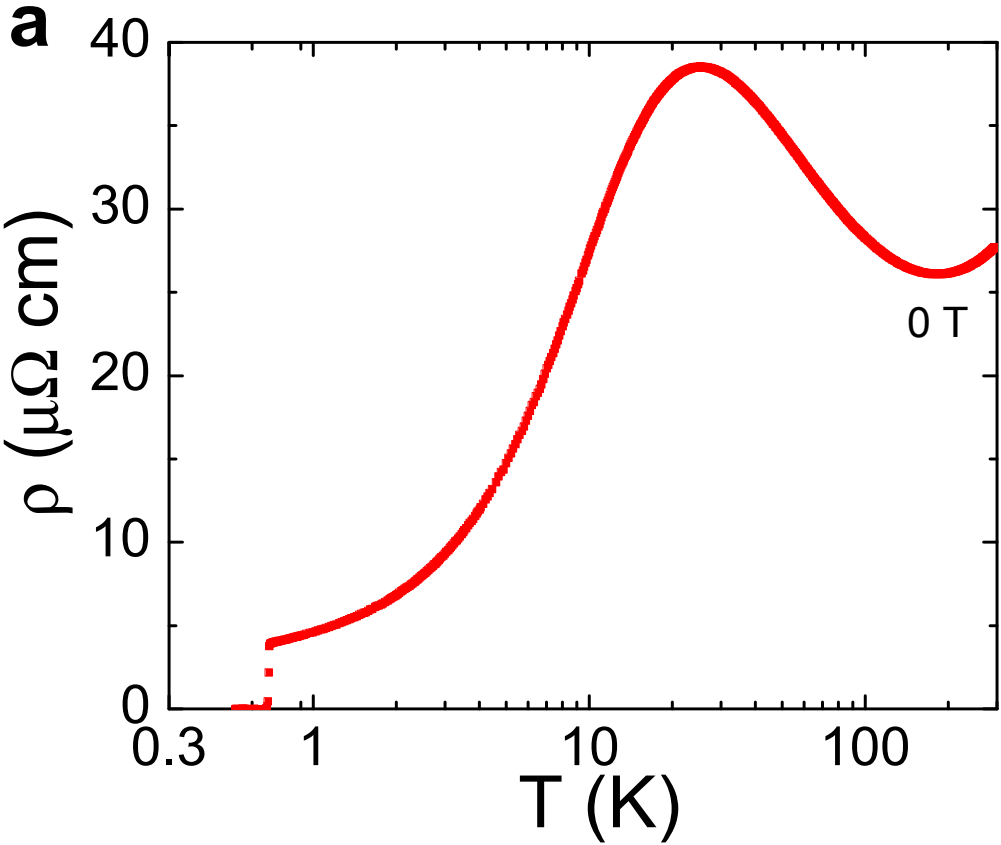
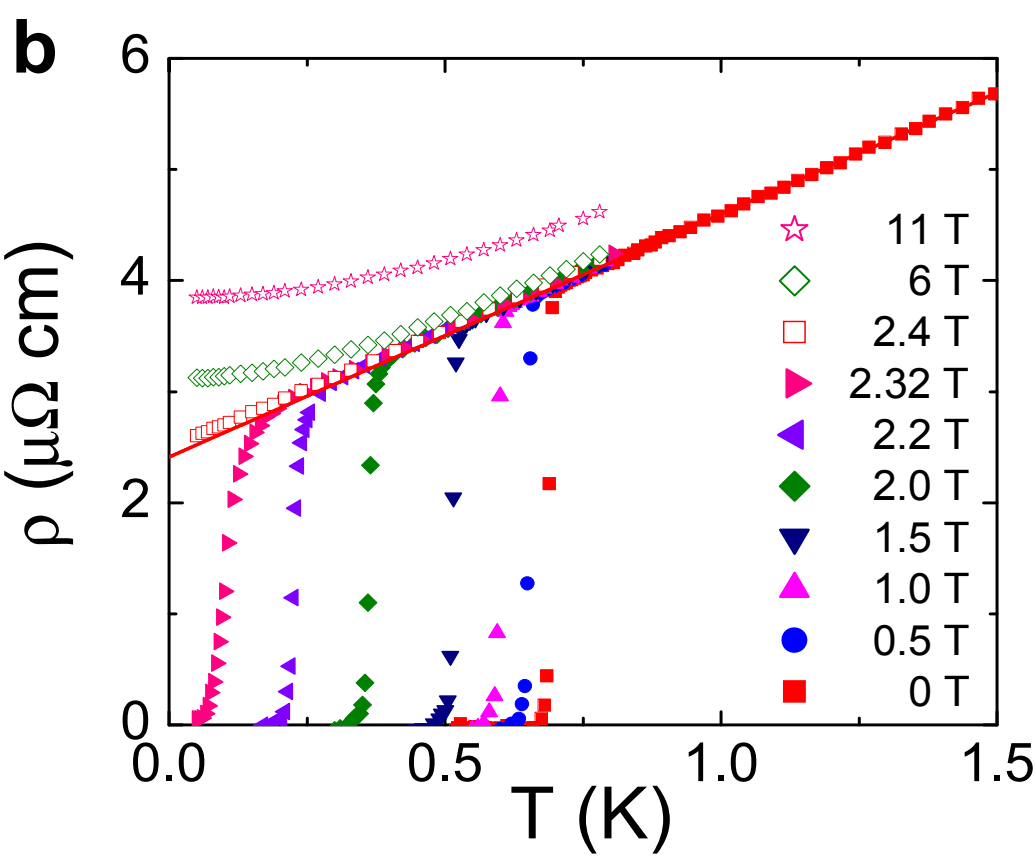
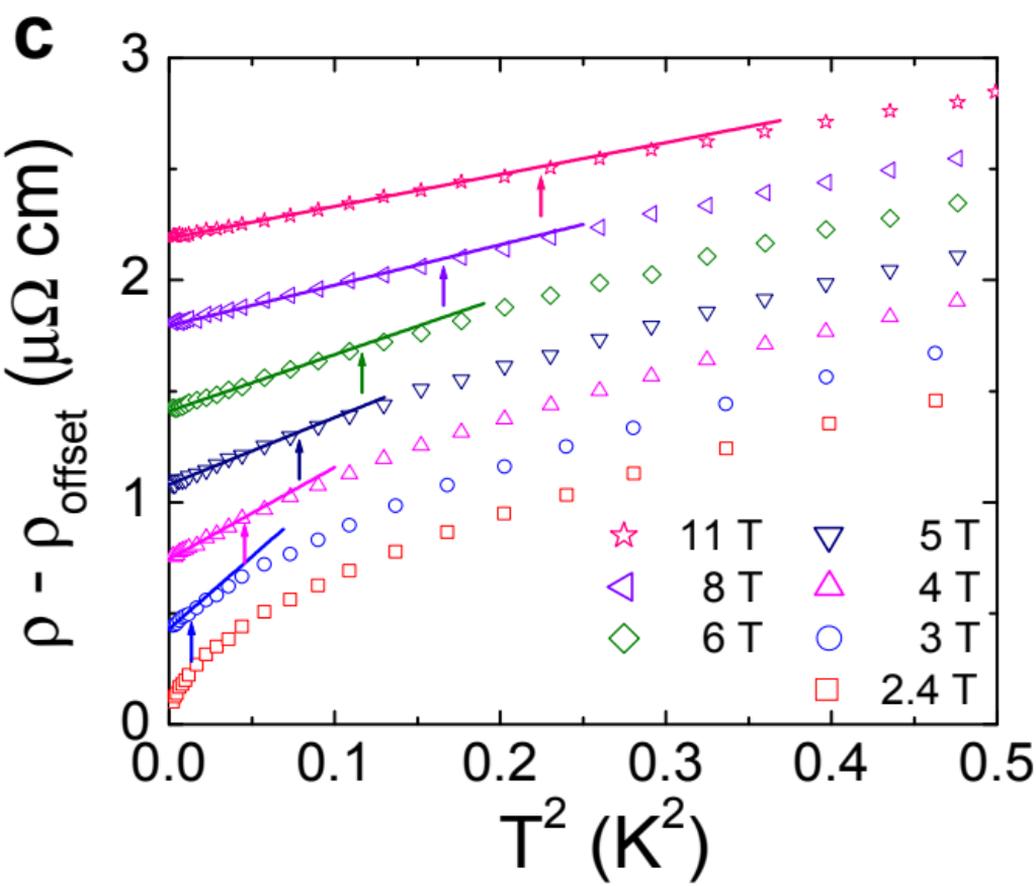

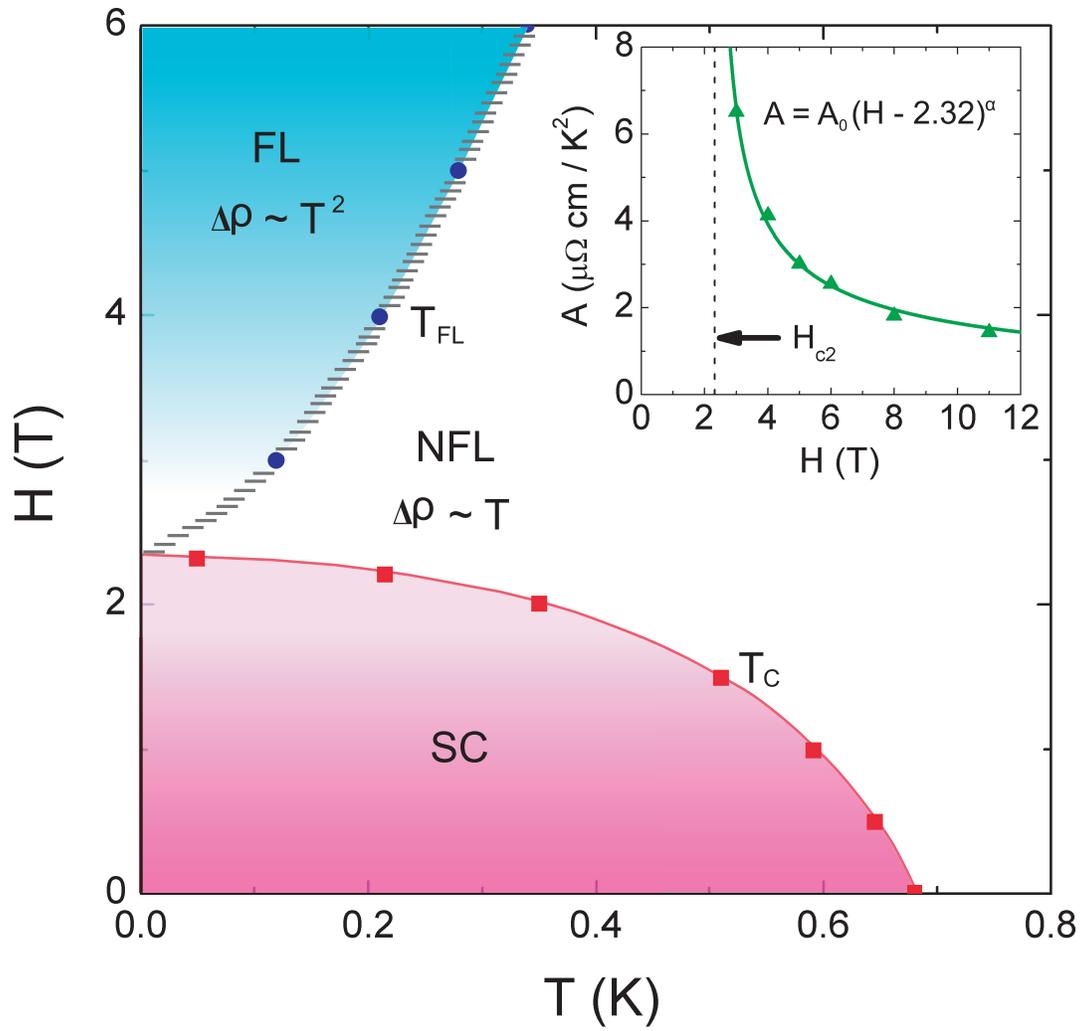

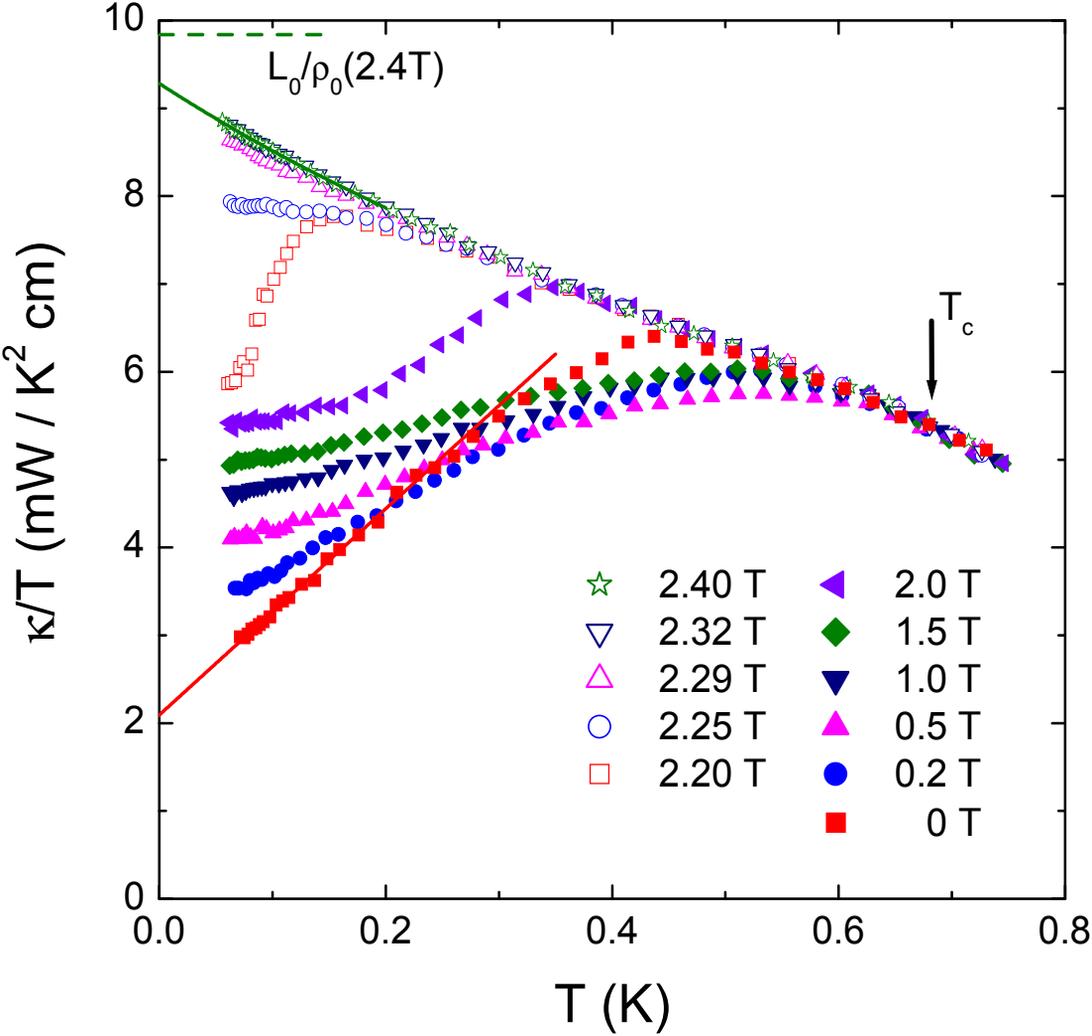

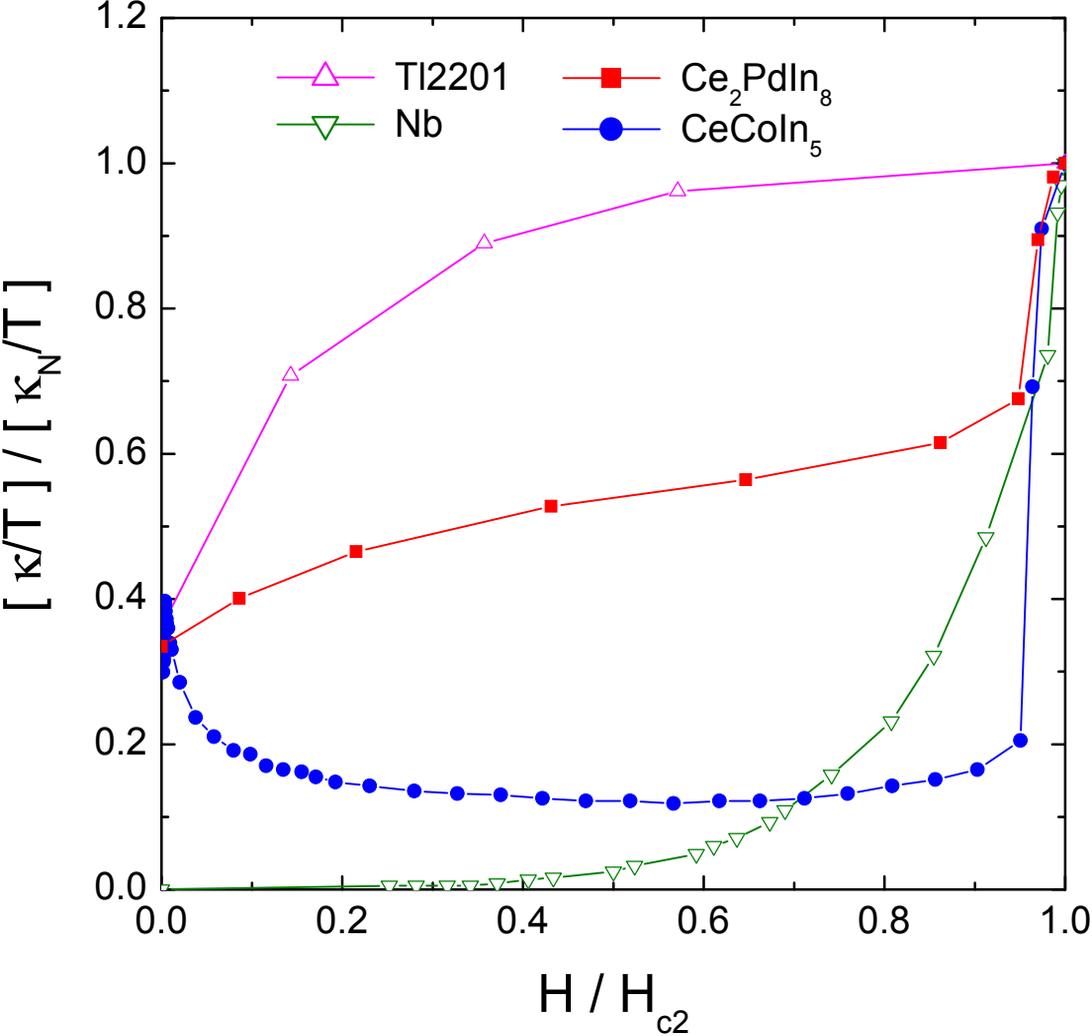